\documentclass[12pt]{iopart}
\usepackage{graphicx}
\usepackage{iopams}
\expandafter\let\csname equation*\endcsname\relax
\expandafter\let\csname endequation*\endcsname\relax
\usepackage{amssymb}
\usepackage{amsmath}
\usepackage{multirow}
\usepackage{xcolor}

\begin{document}

\title{Possible superconductivity in very thin magnesium films}

\author{Giovanni Alberto Ummarino$^{1,2}$ and Alessio Zaccone$^{3,4}$}
\address{$^1$ Istituto di Ingegneria e Fisica dei Materiali,
Dipartimento di Scienza Applicata e Tecnologia, Politecnico di
Torino, Corso Duca degli Abruzzi 24, 10129 Torino, Italy. ORCID 0000-0002-6226-8518}
\address{$^2$ National Research Nuclear University MEPhI (Moscow Engineering
Physics Institute), Kashirskoe shosse 31, Moscow 15409, Russia}
\address{$^3$ Department of Physics ``A. Pontremoli'', University of Milan, via Celoria 16,
20133 Milan, Italy}
\address{$^4$ Institut f{\"u}r Theoretische Physik, University of G{\"o}ttingen,
Friedrich-Hund-Platz 1,
37077 G{\"o}ttingen, Germany}
\ead{giovanni.ummarino@polito.it}
\ead{alessio.zaccone@unimi.it}

\begin{abstract}
It is known that noble metals such as gold, silver and copper are not superconductors, so as magnesium. This is due to the weakness of the electron-phonon interaction which makes them excellent conductors but not superconductors. As it has recently been shown for gold, silver and copper, even for magnesium it is possible that in very particular situations superconductivity may occur.
Quantum confinement in thin films has been consistently shown to induce a significant enhancement of the superconducting critical temperature in several superconductors. It is therefore an important fundamental question whether ultra-thin film confinement may induce observable superconductivity in non-superconducting metals such as magnesium.
We study this problem using a generalization, in the Eliashberg framework, of a BCS theory of superconductivity in good metals under thin-film confinement. By numerically solving these new Eliashberg-type equations, we find the dependence of the superconducting critical temperature on the film thickness, $L$. This parameter-free theory predicts superconductivity in very thin magnesium films. We demonstrate that this is a fine-tuning problem where the thickness must assume a very precise value, close to half a nanometer.
\end{abstract}

\maketitle

\section{Introduction}

It has been seen that thin films of superconductors such as Pb and Al for extremely small thicknesses can produce critical temperatures considerably higher than in the bulk \cite{nostro} thanks to the phenomenon of quantum confinement \cite{zacbcs,nostro}.
We have already seen that in this way it is possible (theoretically) to make even noble metals in the form of very thin films become superconductors \cite{noblemetal}. This happens because the quantum confinement \cite{tony} produces an increase in the electron-phonon interaction thanks to a larger density of states at the Fermi level.
In this article we want to demonstrate that the same theory we used previously predicts that magnesium, in the form of a very thin film, also becomes superconductor.
The magnesium it will transition into the superconducting state at a temperature that is certainly not high but still measurable.
The superconducting critical temperatures will still be low but not so low that they cannot be measured experimentally.
All superconductive physical properties of old low temperature phononic superconductors can be explained in the frameworw of standard one-infinite-band s-wave Eliashberg theory \cite{ummarinorev,revcarbi} essentially in the case of bulk superconductivity.
All the properties of the material, in this theory, are summarized in the spectral function of electron-phonon interaction $\alpha^{2}F(\Omega)$ and in the Coulomb pseudopotential $\mu^{*}$. Once these two quantities are known, it is possible to derive any observable of the material relating to the superconducting state. If one wants to consider the situation of extremely thin films, it is necessary to modify the standard Eliashberg theory.

When the system is no longer bulk but one dimension becomes almost negligible compared to the others (as happens for example in thin films) new phenomena related to quantum confinement come into play. In the context of BCS theory Travaglino and Zaccone \cite{zacbcs} have developed an analytical model that takes into account the thickness of the film. In this model it is possible to reproduce the trend of the critical temperature as a function of the film thickness as has recently been observed experimentally in lead and aluminium films. This theory is connected with the change of the Fermi surface which shows a topological transition in shape when a critical thickness $L_{c}=(2\pi/n(0))^{1/3}$ is reached (here $n(0)$ is the concentration of free carriers). This situation leads to a substantial modification of the electronic density of states which, in the vicinity of the critical thickness, increases substantially, and furthermore it is no longer approximable by a constant around the Fermi level. This approximation is fundamental to write the Eliashberg equations in their simplest standard version.
For example, for Pb thin films it was found \cite{zacbcs} that $L_{c} \approx 4$ {\AA} and this means that this effect, for metal with high density of carriers, appears just in very thin films case.
The theory od Travaglino and Zaccone, as we said, is written
in BCS formalism where some input parameters relating to the material do not have a clear and immediate physical interpretation so it is better
to generalize this theory in the framework of the Eliashberg theory, which is what we will show in the next Section.

\section{Model}
In the Eliashberg theory the material's physical features are taken into account via the electron-phonon spectral function $\alpha^2 F(\Omega)$ and the Coulomb pseudopotential $\mu^{*}$.
These two quantities can be either determined experimentally or calculated from ab initio methods, especially for simple metals. In the simplest version of this theory (one isotropic order parameter and infinite bandwidth) only two of the four terms of self-energy appear: the renormalization function $Z(i\omega_n)$ and the gap function $\Delta(i\omega_n)$ \cite{revcarbi, ummarinorev}. If Migdal's theorem \cite{ummaMig} is satisfied the equations have the following mathematical expression:
\begin{equation}
Z(i\omega_n)=1+\frac{\pi T}{\omega_n}\sum_{\omega_{n'}}\lambda (i\omega_{n'}-i\omega_n)\frac{\omega_{n'}}{\sqrt{\omega_{n'}^2+\Delta^{2}(i\omega_{n'})}}
\end{equation}
\begin{equation}
\Delta(i\omega_n)Z(i\omega_n)=\pi T\sum_{\omega_{n'}}\big[ \lambda (i\omega_{n'}-i\omega_n)-\mu^{*}(\omega_{c})\theta(\omega_{c}-|\omega_{n'}|)\big]\\
\frac{\Delta(i\omega_{n'})}{\sqrt{\omega_{n'}^2+\Delta^{2}(i\omega_{n'})}}\\
\end{equation}
 $n$ are integer numbers related to Matsubara energies $\omega_{n}$ , $\mu^{*}(\omega_{c})$ is the Coulomb pseudopotential that depends, in a weak way, on a cut-off energy $\omega_{c}$ ($\omega_{c}> 3\Omega_{max}$ where $\Omega_{max}$ is the maximum phonon energy) \cite{Allen}, and $\theta(\omega_{c}-|\omega_{n'}|)$ is the Heaviside function.
The electron-phonon spectral function $\alpha^2F(\Omega)$ is present inside $\lambda (i\omega_{n'}-i\omega_n)$ is this way:

\begin{equation}
\lambda (i\omega_{n'}-i\omega_n)=2\int_0^\infty \frac{\Omega\alpha^2F(\Omega) d\Omega}{\Omega^2+(\omega_{n'}-\omega_n)^2}.
\end{equation}

The strength of the electron-phonon coupling is given by the electron-phonon coupling parameter $\lambda=2\int_0^\infty \frac{\alpha^2F(\Omega) d\Omega}{\Omega}$.
In general, it is impossible to find exact analytical solutions of Eliashberg's equations except for the case of extreme strong-coupling ($\lambda>10$) \cite{revcarbi}. Hence, we solve them numerically with an iterative method until numerical convergence is reached. We have shown the theory in the formulation on the imaginary axis because the numerical solution is more easy to find but it exists also in the version on the real axis.
In principle the critical temperature can be calculated by solving an eigenvalue equation but it is more simple by giving a very small test value to the superconducting gap (for example $10^{-10}$ times the value at zero temperature) and then by checking at which temperature the solution converges. In this way, it is possible to obtain a precision in the $T_{c}$ which is much larger than any possible experimental verification.
When we takes into account effects of quantum confinement on the free carriers it is necessary to modify the Eliashberg theory and the equations will be written in a more complex shape as well as increasing in number \cite{ummag,ummachi}. The effect of confinement appears, essentially, in the normal density of states ($NDOS$) around the Fermi level cannot be approximated by a constant value). We have already shown, in a previous article, that this new theory \cite{nostro,noblemetal}, devoid of free parameters explain the increase of critical temperature \cite{nostro} in the very thin films of $Pb$ and $Al$ as well ass it predicts superconductivity for $Au$, $Ag$ and $Cu$ ultra thin films \cite{noblemetal}. In this way the noble metals become superconductors at precise values of the film thickness $L$.
If the $NDOS$ is not more a constant  but a function of energy, the Eliashberg equations become slightly more complex and they become four equations \cite{Allen}.
An another step in the generalization of the theory is to remove the infinite band approximation (which works very well for most metals in the bulk state) \cite{ummag,ummachi}.
In the more general situation we have four equations to solve\cite{ummachi} but in the particular case where the normal density of states is symmetrical with respect to the Fermi level ($N(\varepsilon)=N(-\varepsilon)$), it is possible to simplify the theory in a way that the self energy terms remain just two, $Z(i\omega_n)$ and $\Delta(i\omega_n)Z(i\omega_n)$ as before, and the new equations read as \cite{carbin1}
\begin{equation}
\begin{split}
Z(i\omega_n)= 1+\frac{\pi T}{\omega_n}\sum_{\omega_{n'}}\lambda (i\omega_{n'}-i\omega_n)\frac{2}{\pi}\arctan(\frac{W}{2Z(i\omega_{n'})\sqrt{\omega_{n'}^{2}+\Delta^{2}(i\omega_{n'})}}) \times
\\
\times \frac{\omega_{n'}}{\sqrt{\omega_{n'}^2+\Delta^{2}(i\omega_{n'})}}[\frac{N(i\omega_{n'})+N(-i\omega_{n'})}{2}]
\end{split}
\end{equation}

\begin{equation}
\begin{split}
\Delta(i\omega_n)Z(i\omega_n)=\pi T\sum_{\omega_{n'}}\big[\lambda(i\omega_{n'}-i\omega_n)-\mu^{*}(\omega_c)\theta(\omega_c-|\omega_{n'}|)\big]\frac{2}{\pi} \arctan(\frac{W}{2Z(i\omega_{n'})\sqrt{\omega_{n'}^{2}+\Delta^{2}(i\omega_{n'})}})\times
\\
\times \frac{\Delta(i\omega_{n'})}{\sqrt{\omega_{n'}^2+\Delta^{2}(i\omega_{n'})}}[\frac{N(i\omega_{n'})+N(-i\omega_{n'})}{2}]
\end{split}
\end{equation}
where $N(\pm i\omega_{n'})=N(\pm Z(i\omega_{n'})\sqrt{(\omega_{n'})^{2}+\Delta^{2}(i\omega_{n'})})$ and the bandwidth $W$ is equal to half the Fermi energy, $E_{F}/2$.
The fact that normal density of states symmetric with respect to Fermi level is a great advantage and allows to find the numerical solution more quickly.
When the effects of quantum confinement begin to manifest themselves, the NDOS can no longer be approximated by its value at the Fermi level and two different regimes can appear\cite{zacbcs}: $L>L_{c}$ and $L<L_{c}$.

If we are in the case $L>L_{c}$ and, consequently, $E_{F}>\varepsilon^{*}$), the normal density of states has the following form
\begin{equation}
N(\varepsilon)=N(0)C
[\theta(\varepsilon^{*}-\varepsilon)\sqrt{\frac{E_{F}}{\varepsilon^{*}}}\frac{|\varepsilon|}{E_{F}}+
\theta(\varepsilon-\varepsilon^{*})\sqrt{\frac{|\varepsilon|}{E_{F}}}]
\end{equation}
where $C=(1+\frac{1}{3}\frac{L_{c}^{3}}{L^{3}})^{1/3}$, $\varepsilon^{*}=\frac{2\pi^{2}\hbar^{2}}{mL^{2}}$, $L_{c}=(\frac{2\pi}{n_{0}})^{1/3}$, $m$ is the electron mass and $E_{F,bulk}$ is the Fermi energy of the bulk material. In this case, it is possible to demonstrate the following relations \cite{zacbcs}:

\begin{equation}
E_{F}=C^{2}E_{F,bulk}
\end{equation}

\begin{equation}
N(E_{F})=C N(E_{F,bulk})=CN(0)
\end{equation}

with
\begin{equation}
N(E_{F,bulk})=\frac{V(2m)^{3/2}}{2\pi^{2}\hbar^{3}}\sqrt{E_{F,bulk}}.
\end{equation}

In the regime $\epsilon < \epsilon^*$, the $NDOS$ has a new, linear dependence on the energy, in contrast with the standard square-root dependence which is retrieved for $\epsilon > \epsilon^*$ \cite{zacbcs}.
In order to better understand the new physics hidden in these equations
we remove the factor $C$, which will be put in the renormalization of the electron-phonon coupling constant.

We now recap all the changes that are present in the new Eliashberg theory:
i) the normal density of states will no longer be a constant but a function of energy;

\begin{equation}
N(\varepsilon)=\left[\vartheta(\varepsilon^{*}-\varepsilon)\sqrt{\frac{E_{F}}{\varepsilon^{*}}}\frac{|\varepsilon|}{E_{F}}+
\vartheta(\varepsilon-\varepsilon^{*})\sqrt{\frac{|\varepsilon|}{E_{F}}}\right]
\end{equation}

ii) the electron-phonon interaction is a function of film thickness $L$, via $\lambda=C\lambda^{bulk}$ because we move the prefactor of the $NDOS$, $C$, inside the definition of electron-phonon coupling as in the Coulomb pseudopotential. This choice allows one to justify the use of Allen-Dynes equation \cite{Dynes} for $T_{c}$. Of course the shape of electron-phonon spectral function remains the same and just we rescale it for changing the value of electron-phonon coupling constant;\\
iii) the value of the Fermi energy is also a function of the film thickness $L$: $E_{F}=C^{2}E_{F,bulk}$. Of course, in the symmetric case discussed above, it is $W=E_{F}/2$;\\
iv) the Coulomb pseudopotential $\mu^{*}$ also depends on the film thickness via
\begin{equation}
\mu^{*}=\frac{C\mu_{bulk}}{1+\mu_{bulk}\ln(E_{F}/\omega_{c})}
\end{equation}
where $\mu_{bulk}=\frac{\mu^{*}_{bulk}}{1-\mu^{*}_{bulk}\ln(E_{F,bulk}/\omega_{c})}$.\\
Instead, when $L<L_{c}$ and, consequently, $E_{F}<\varepsilon^{*}$, we have \cite{zacbcs}:
\begin{equation}
N(\varepsilon)=C'N(0)\sqrt{\frac{E_{F}}{\varepsilon^{*}}}\frac{\varepsilon}{E_{F}}
\end{equation}

\begin{equation}
N(\varepsilon=E_{F})=C'N(0)
\end{equation}

\begin{equation}
E_{F}=C'^{2}E_{F,bulk}\\
\end{equation}
where
\begin{equation}
C'=\frac{2}{6^{1/3}}\sqrt{\frac{L}{L_{c}}}
\end{equation}

Now, the normal density of states is given by \cite{zacbcs}:
\begin{equation}
N(\varepsilon)=\sqrt{\frac{E_{F}}{\varepsilon^{*}}}\frac{|\varepsilon|}{E_{F}}
\end{equation}
The electron-phonon coupling and the Coulomb pseudopotential take, through $C'$, a dependence from the thickness:

\begin{equation}
\lambda=C'\lambda^{bulk} ,~~~ \mu^{*}=\frac{C'\mu_{bulk}}{1+\mu_{bulk}\ln(E_{F}/\omega_{c})}.
\end{equation}

In the standard Eliashberg equations the reference energy is the Fermi energy which is represents the zero of energy.\par
In the code that numerically solves the Eliashberg equations, by recalling that the reference energy is the Fermi energy taken as the zero of the energy, the normal density of states has been rescaled in the following way
(of course the normal density of states will be continuous for $\varepsilon=\varepsilon^{*}$). When $L>L_{c}$ and $\varepsilon^{*}<E_{F}$:

\begin{equation}
N(\varepsilon)=[\vartheta(\varepsilon^{*}-\varepsilon)\sqrt{\frac{E_{F}}{E_{F}-\varepsilon^{*}}}(1-\frac{|\varepsilon|}{E_{F}})+
\vartheta(\varepsilon-\varepsilon^{*})(1-\sqrt{\frac{|\varepsilon|}{E_{F}}})].
\end{equation}

Instead, when $L<L_{c}$ and $\varepsilon^{*}>E_{F}$:

\begin{equation}
N(\varepsilon)=\sqrt{\frac{E_{F}}{\varepsilon^{*}}}(1-\frac{|\varepsilon|}{E_{F}}).
\end{equation}
\begin{figure*}[ht]
	\centerline{\includegraphics[width=1\textwidth]{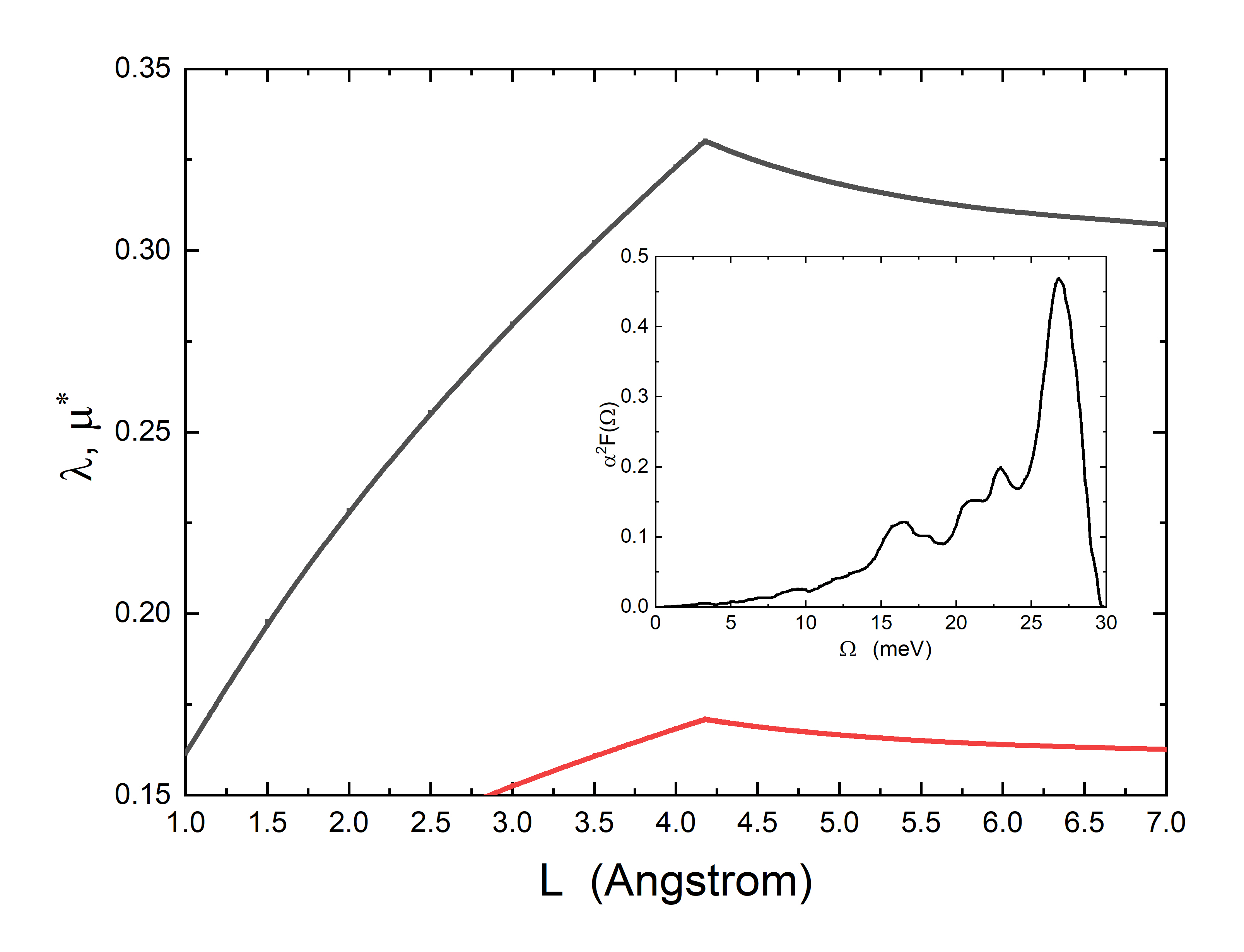}}
\caption{Physical parameters used in the theory for magnesium films: $\lambda$ (full black solid line) and $\mu^{*}$ (full red solid line). All parameters are plotted as a function of the film thickness $L$.  In the inset, the Eliashberg electron-phonon spectral function of magnesium is shown, from Ref. \cite{a2fMg}.}\label{diagrams1}
\end{figure*}

\begin{figure*}[t!]
	\centerline{\includegraphics[width=1\textwidth]{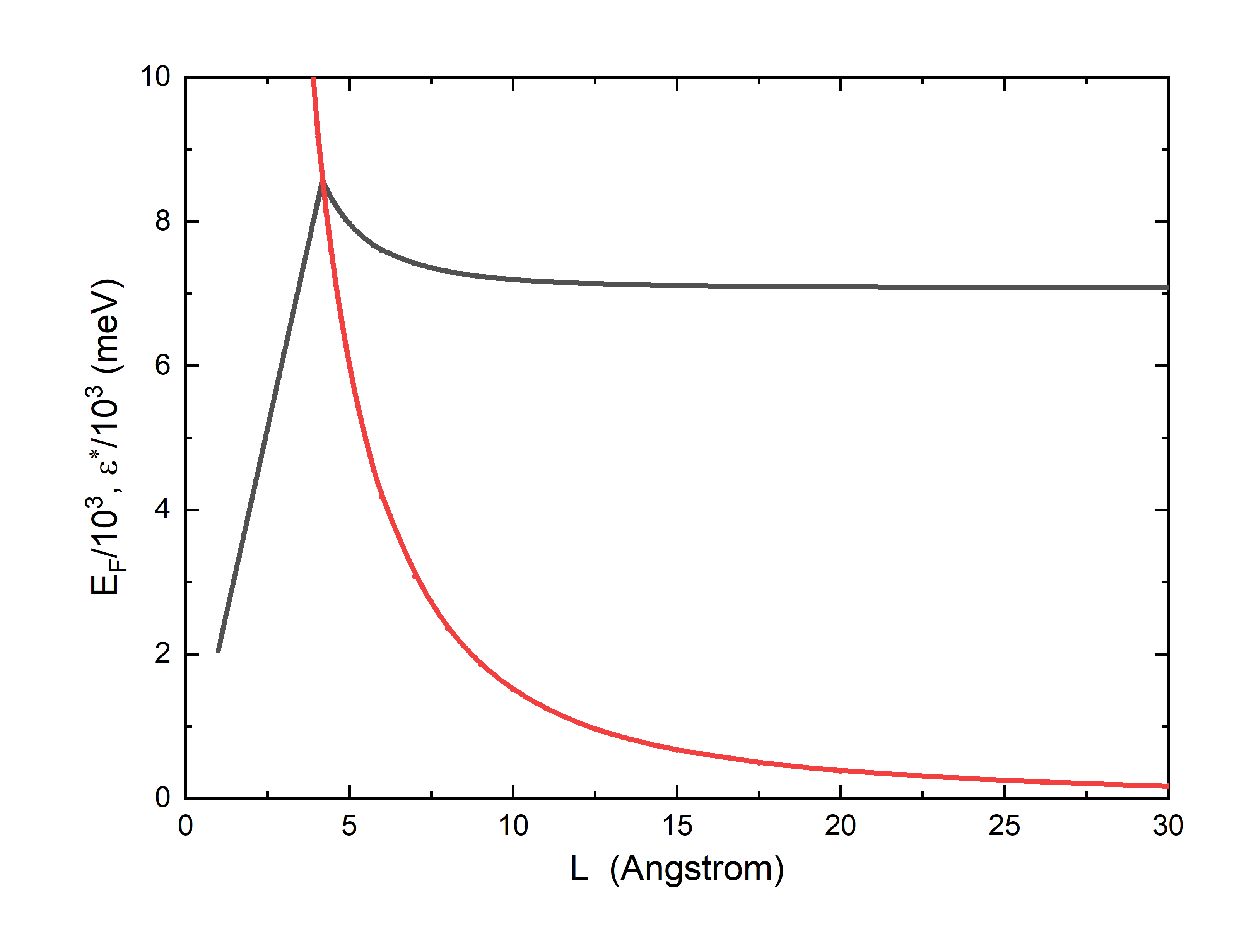}}
\caption{Physical parameters used in the theory for magnesium films: $\varepsilon^{*}/5\cdot 10^{3}$) (full red solid line) and $E_{F}/\cdot 10^{3}$ (full black solid line). All parameters are plotted as a function of the film thickness $L$.}\label{diagrams2}
\end{figure*}

\begin{figure*}[t!]	\centerline{\includegraphics[width=1\textwidth]{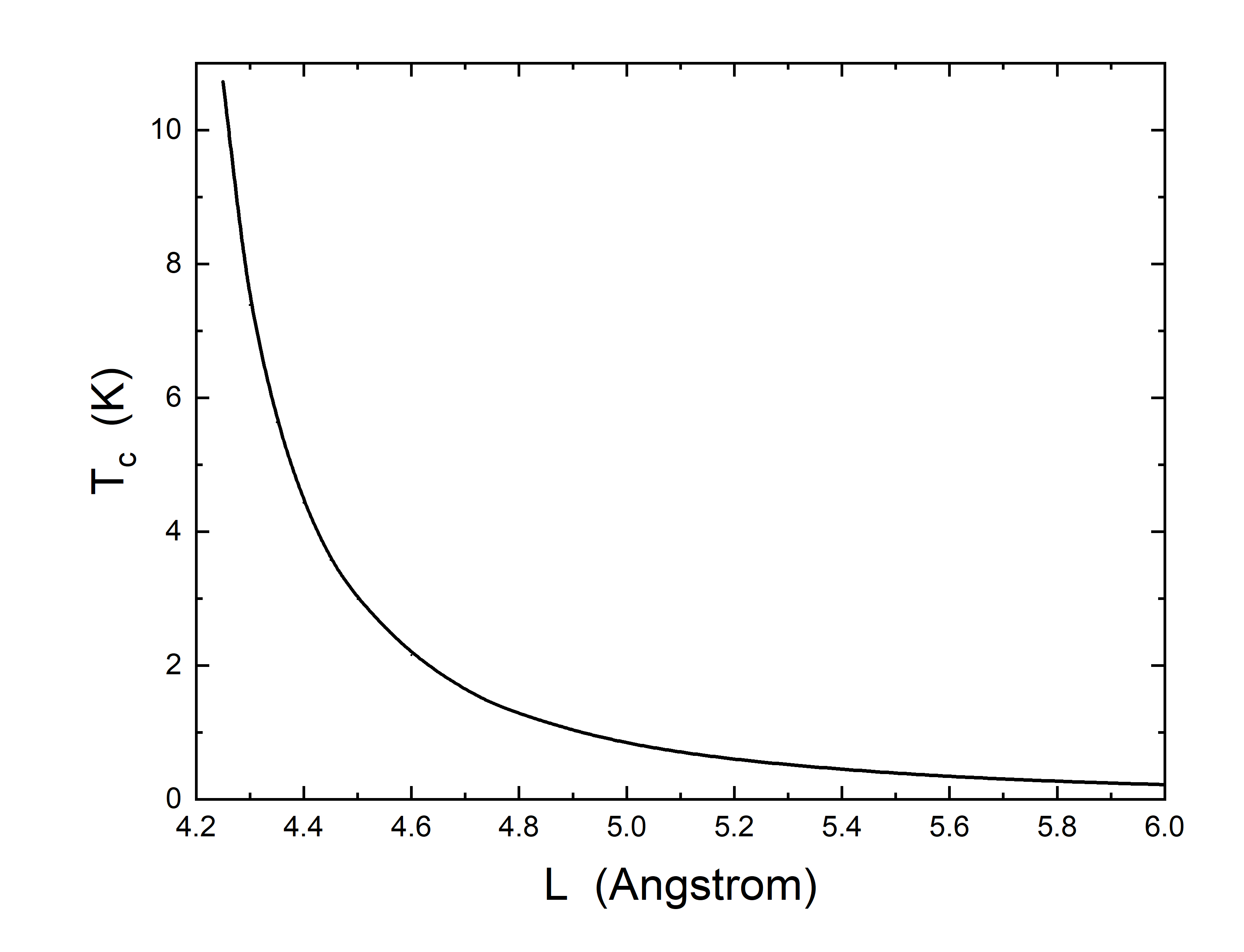}}
\caption{Critical temperature $T_c$ versus film thickness $L$: full solid line represent the numerical solutions of Eliashberg equations.}\label{diagrams3}
\end{figure*}

\begin{figure*}[t!]	\centerline{\includegraphics[width=1\textwidth]{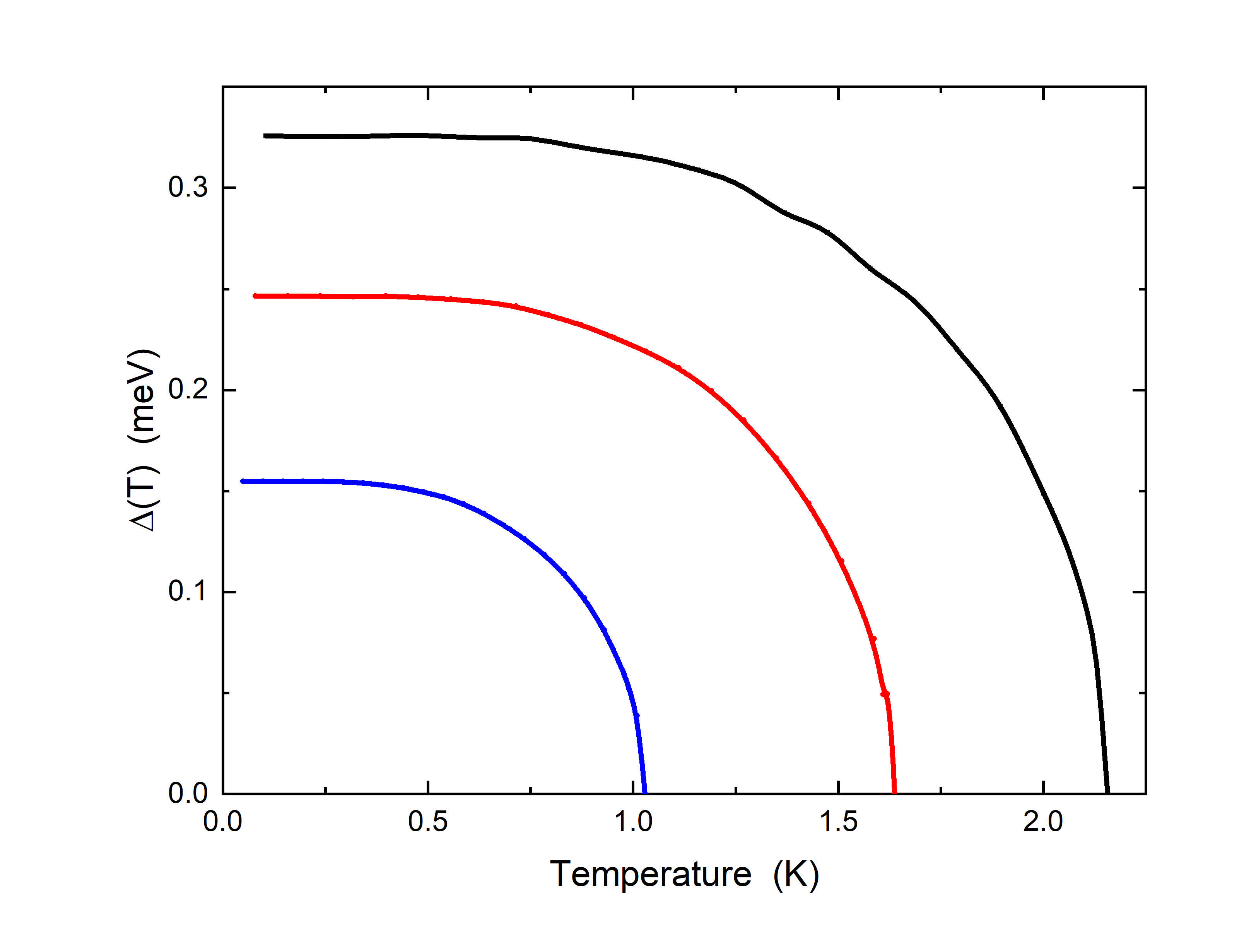}}
\caption{Superconductive gap $\Delta$ versus temperature for three different film thickness $L=4.6, 4.7, 4.9$ {\AA}: full solid line black, red and dark blue represent the numerical solutions of Eliashberg equations.}\label{diagrams3}
\end{figure*}
\section{Prediction on critical temperature}
We have seen that, if the $NDOS$ is symmetrical, the theory is simplified and we have just two Eliashberg equations to solve. Instead the theory becomes more complex if the $NDOS$ is aymmetrical and the equations to be solved are three \cite{ummachi}. It is important underline that the relevant fact is the non constant $NDOS$ more than the symmetry of the same $NDOS$. Usually the asymmetry is a problem of second order and can become significant only in very particular situations \cite{carbin1}.
This theory is completely general and can be easily extended to multiband metals \cite{UmmaMulti1,UmmaMulti2,UmmaMulti3}.
As we have demonstrated in a previous article, noble metals ($Au$, $Ag$, $Cu$), although they have a very weak electron-phonon coupling ($\lambda<0.25$) can be superconductors if
are in the shape of very thin films \cite{noblemetal} with a thickness very close to the critical length $L_{c}$, which is of the order of 5 {\AA} (0.5 nm).
This happen because the electron-phonon interaction is greatly enhanced and in a narrow range of thickness to produce superconductivity at experimentally accessible temperatures.
The same thing happens for magnesium as is revealed by our calculations in Figs. 1-3.

In Fig. 1 two physical quantities of magnesium, the electron phonon coupling constant and the Coulomb pseudopotential, used in the theoretical calculations are plotted as functions of the film thickness $L$. The bulk electron-phonon spectral function of magnesium \cite{a2fMg} with $\lambda_{bulk}=0.30$ is shown in the inset of Fig. 1.
The cut-off energy is $\omega_{c}=90$ meV and is related to bulk value of the Coulomb pseudopotential \cite{tunnMg} $\mu^{*}(\omega_{c})=0.16$ while the maximum electronic energy is $\omega_{max}=150$ meV. The values of the bulk Fermi energy and carrier density are, respectively, $E_{F,bulk}=7080$ meV and $n_{0}=0.0861\cdot 10^{30}$ $m^{-3}$ \cite{Mermin}. This produces a critical thickness $L_{c}=4.18$ {\AA}. In Fig. 2 are shown the other two physical quantities that are present in the theory: the thickness $L$.
In Fig. 1(a) it is shown what we anticipated in the text of the paper and precisely, that, around the critical thickness value, the coupling constant $\lambda$ has a slight increase. Will this increase in the value of the electron-phonon coupling constant be sufficient to produce the superconducting state? To check this we will solve the modified Eliashberg equations and calculate the critical temperature $T_c$. The result is shown in Fig. 3.
We find that, for the film thickness $L=4.40$ {\AA} (very close to the critical value $L_{c}=4.18$ {\AA}) the material becomes a superconductor with $T_{c}=4.43$ $K$. We notice that the thickness range that allows superconductivity to exist is quite narrow, which can be understood based on the underlying topological-type transition \cite{zacbcs}.
We can see that, as soon as we move away from the critical value $L_c$ of film thickness, the $T_c$ abruptly goes to very small values, which we are not able to calculate as it is too time-consuming for the code to reach convergence.
Finally, we should also point out that solid films that are as thin as $0.5$ nm are still effectively described by three-dimensional physics as shown plenty of times in the literature on the basis of experiments, theory and atomistic simulations, e.g. cfr. \cite{zacbcs,Yu}, albeit with substantial corrections due to confinement such as those implemented in our theory.
\section{Conclusions}

For including the crucial effect of quantum confinement we have generalized the Eliashberg theory where the thickness of the thin film appears as well as the free carriers density
In this way we are able to compute the superconducting properties of magnesium thin films, in a fully quantitative way and with no free parameters. Upon decreasing the film thickness, the Fermi surface shape changes
as well as the $NDOS$ and this fact leads to the increase in the number of electronic states at the Fermi level. This situation bring to increase, significantly, the electron-phonon coupling, and hence to have, surprisingly superconductivity to low but experimentally measurable temperatures. These theoretical predictions reveal the possibility that magnesium thin films with a thickness close to 0.4-0.5 nm become superconducting. These predictions are relevant from both a fundamental and an applied point of view.
\subsection{Acknowledgments}
A.Z. gratefully acknowledges funding from the European Union through Horizon Europe ERC Grant number: 101043968 ``Multimech'', from US Army Research Office through contract nr. W911NF-22-2-0256, and from the Nieders{\"a}chsische Akademie der Wissenschaften zu G{\"o}ttingen in the frame of the Gauss Professorship program.

\newpage


\begin{thebibliography}
\small{
%
\bibitem{nostro} Giovanni Alberto Ummarino, Alessio Zaccone, \textit{J. Phys.: Condens. Matter} \textbf{37}, 065703 (2025).
%
\bibitem{zacbcs} R. Travaglino and A. Zaccone, \textit{Journal of Applied Physics} \textbf{133}, 033901 (2023).
%
\bibitem{noblemetal} Giovanni Alberto Ummarino, Alessio Zaccone, \textit{Phys. Rev. B} \textbf{8}, L101801 (2024).
%
\bibitem{tony} A. Bianconi and M. Missori, \textit{J. Phys. I} \textbf{4} 4, 361 (1994).
%
\bibitem{ummarinorev} G.A. Ummarino, in \textit{Emergent Phenomena in Correlated
Matter}, edited by E. Pavarini, E. Koch, and U. Schollwöck
(Forschungszentrum Jülich GmbH and Institute for Advanced
Simulations, Jülich, Germany, 2013), pp. 13.1–13.36
%
\bibitem{revcarbi} J.P. Carbotte, \textit{Rev Modern Physics} \textbf{62}, 61 (1990).
  %
\bibitem{ummaMig} G.A. Ummarino, R.S. Gonnelli, \textit{Physical Review B} \textbf{56}, R14279, (1997).
%
\bibitem{ummag} G.A. Ummarino, R.S. Gonnelli, V.A. Stepanov, \textit{Nuovo Cimento della Societa Italiana di Fisica D } \textbf{19}, 1215, (1997).
%
\bibitem{ummachi} G.A. Ummarino, R.S. Gonnelli, \textit{Physical Review B} \textbf{66}, 104514, (2002).
%
\bibitem{Allen} P.B. Allen, \textit{Phys. Rev. B} \textbf{17} 3725-3734 (1978).
%
\bibitem{carbi1n} B. Mitrovi´c and J. Carbotte, \textit{Can. J. Phys.} \textbf{61} 784 (1983).
%
\bibitem{a2fMg} A. Leonardo, I.Yu. Sklyadneva, V.M. Silkin, P.M. Echenique, and E.V. Chulkov, \textit{Phys. Rev. B} \textbf{76}, 035404 (2007).
%
\bibitem{tunnMg} D.M. Burnell  and E.L. Wolf, \textit{Journal of Low Temperature Physics} \textbf{58}, 1027 (1984).
%
\bibitem{Mermin} Neil Ashcroft, David Mermin,"Solid State Physics", Cengage Learning, Inc (2021).
%
\bibitem{Dynes} P.B. Allen and R.C. Dynes, \textit{Phys. Rev. B} \textbf{12}, 905 (1975).
%
\bibitem{carbin1} E.Schachinger and J.P. Carbotte, \textit{J. Phys. F: Met. Phys.} \textbf{13}, 2615 (1983).
%
\bibitem{UmmaMulti1} D. Torsello, G.A, Ummarino, L. Gozzelino, T. Tamegai, G. Ghigo,
\textit{Phys. Rev. B} \textbf{9}, 134518 (2019).
%
\bibitem{UmmaMulti2} D. Torsello, G.A. Ummarino, J. Bekaert, L. Gozzelino, R. Gerbaldo,
M.A. Tanatar, P.C. Canfield, R. Prozorov, and G. Ghigo
\textit{Physical Review Applied} \textbf{13}, 064046 (2020).
%
\bibitem{UmmaMulti3} G. Ghigo, G.A. Ummarino, L. Gozzelino, T. Tamegai,
\textit{Phys. Rev. B} \textbf{96}, 014501 (2017).
%
\bibitem{UmmaMulti4} D. Torsello, K. Cho, K.R. Joshi, S. Ghimire, G.A. Ummarino, N.M. Nusran,
M.A. Tanatar, W. R. Meier, M. Xu, S.L. Bud’ko, P.C. Canfield, G. Ghigo, and R. Prozorov,
\textit{Phys. Rev. B} \textbf{100},  094513 (2019).
%
\bibitem{Yu} Y. Yu, C. Yang, M. Baggioli, Anthony E. Phillips, Alessio Zaccone,
Lei Zhang, Ryoichi Kajimoto, Mitsutaka Nakamura, Dehong Yu, Liang Hong, \textit{Nat. Commun.} \textbf{13}, 3649 (2022).
}
\end{thebibliography}
\end{document}